\documentstyle[preprint,aps]{revtex}
\begin{document}
\draft

\title{Series Expansions for
Excited States of Quantum Lattice Models}

\author{Martin P. Gelfand}
\address{Department of Physics, Colorado State University,
Fort Collins, Colorado 80523}

\maketitle

\begin{abstract}

We show that by means of connected-graph expansions
one can effectively generate exact high-order series expansions
which are informative of low-lying excited states for
quantum many-body systems defined on a lattice.
In particular, the Fourier series coefficients of elementary
excitation spectra are directly obtained.
The numerical calculations involved are straightforward extensions of
those which have already been used to calculate
series expansions for ground-state correlations and $T=0$
susceptibilities in a wide variety of models.
As a test, we have reproduced the known elementary excitation
spectrum of the transverse-field Ising chain in its disordered phase.

\end{abstract}


\bigskip

Several years ago, Singh, Huse and this author \cite{gsldqaf,peqmbs}
described an efficient, general method for calculating high-order
perturbations expansions for ground-state correlations and $T=0$
susceptibilities in quantum many-body systems defined on a lattice.
(It is worth noting, belatedly, that we rediscovered methods
that had been developed first by Hamer, Nickel, and others;
a convenient entry point into that literature is
He {\it et al.} \cite{Heetal})
This method had two principle facets: (1) A connected-cluster
(a.k.a. ``linked-cluster'', ``connected-graph'', or simply ``cluster'')
expansion
for the properties of interest, combined with (2) a formulation of
Rayleigh-Schr\"odinger perturbation theory in terms of a recurrence
relation, for the purpose of constructing the ground state energy
and eigenvector perturbatively on each cluster.
This method has been applied to a wide variety of problems involving
$S=1/2$ and $S=1$ antiferromagnets \cite{spinpapers} and also to the
Kondo lattice model of mobile fermions and stationary
spins \cite{kondopaper}.
In every case one considers a family of Hamiltonians parameterized
by $\lambda$,
\begin{equation}
H =  H_0 +  \lambda H_1,
\end{equation}
in which the ``unperturbed'' Hamiltonian is trivially diagonalizable
and of fixed degeneracy on any connected cluster of sites.

The perturbation expansion approach is comparable in many respects to
finite-size diagonalization studies.
A major advantage of the former is that the series expansions do not
suffer from finite-size and shape effects --- they are exact
for the infinite lattice.
A major disadvantage has been that
many important properties associated with excited states,
such as elementary excitation spectra and their associated residues,
were inaccessible.
Here I show that these properties can be calculated in
terms of perturbation expansions about trivial unperturbed
Hamiltonians, by direct (but hardly obvious, {\it a priori})
extensions of the two crucial facets of the method for
ground state properties which are listed above.
The algorithms are simple in principle, and if computer code for
generating the ground-state series expansions for some model already
exists then it is straightforward to modify it so as to yield
excited-state data.

The key to the approach is to construct the {\it effective
Hamiltonian\/} within a degenerate (for $H_0$) manifold of
excited states for each finite cluster.
The formulae which yield perturbation expansions for the effective
Hamiltonians are derived by considering a similarity transformation
\begin{equation}
S^{-1} H S  = H^{\rm eff}
\end{equation}
where $H$ is the Hamiltonian restricted to some finite cluster,
and written explicitly in the basis of the eigenstates of $H_0$.
For definiteness
we assume there are $L$ degenerate states out of $M$ states total,
and it is useful to order the eigenstates
of $H_0$ so that the first $L$ are those of the manifold.
We can choose $S$ so as to not mix the states within the degenerate
manifold, that is, we take it to be the identity within the
upper left $L \times L$ block.
The matrix $S$ is to be constructed
so that in $H^{\rm eff}$ there are vanishing matrix elements between
the degenerate manifold and the other states, that is,
$H^{\rm eff}$ breaks up onto an $L \times L$ block in the upper left,
describing the interactions between the ``dressed'' states of
the manifold, and an $(M-L) \times (M-L)$ block, describing
all the other (irrelevant) physics of the cluster.
Traditionally (for example, the Foldy-Wouthuysen transformation
of the Dirac equation), $S$ would be written as $e^{iT}$ and the
Hermitian operator $T$ written as a power series in $\lambda$,
the exponentials would be expanded, and the vanishing of the appropriate
$H^{\rm eff}$ matrix elements would lead to a set of equations
for the terms in the series for $T$.
Rather than expanding exponentials, we construct
$H^{\rm eff}$ simply by moving $S^{-1}$ to the other side and
viewing $S$ as
a set of column vectors $\psi^{(l)}$.
We now have an equation in which the vectors $\psi^{(l)}$ with
$1\leq l \leq L$ are not coupled to those with $L+1 \leq l \leq M$,
and we can now entirely neglect the part of $H^{\rm eff}$
that lies outside the $L \times L$ block.
Then expand everything in sight in powers of $\lambda$:
\begin{equation}
\psi^{(l)}=\sum_k \lambda^k \psi^{(l)}_k,
\end{equation}
\begin{equation}
H^{\rm eff}(l',l)=\sum_k \lambda^k H^{\rm eff}_k(l',l)
\end{equation}
and collect the terms proportional to $\lambda^k$ in column $l$ to yield
\begin{equation}
H_0 \psi_k^{(l)} + H_1 \psi_{k-1}^{(l)} =
   \sum_{k'=0}^k\sum_{l'=1}^L \psi_{k'}^{(l')} H^{\rm eff}_{k-k'}(l',l) .
\end{equation}

{}From this equation one can now almost immediately write coupled recurrence
relations for the terms in the power series expansions of
$H^{\rm eff}$ and the $\psi^{(l)}$.
It is only necessary to
impose a ``normalization convention'' (which affects the form
of the final equations but not any of the physical results):
it is most convenient to require for states $|l\rangle$ in the
degenerate manifold of $H_0$ that
\begin{equation}
\langle l | \psi_k^{(l')} \rangle = \delta_{k,0}\delta_{l,l'} .
\end{equation}
Then one has for the effective Hamiltonian
\begin{equation}
H^{\rm eff}_k(m,l)= \langle m | H_1 | \psi_{k-1}^{(l)} \rangle,
\end{equation}
and for the projection of $| \psi^{(l)} >$ onto an eigenstate state of
$H_0$  outside the manifold the result is
\begin{equation}
\langle n | \psi_k^{(l)} \rangle = {1\over E_0-E_n }
\left(
\langle n | H_1 | \psi_{k-1}^{(l)} \rangle -
\sum_{k'=1}^{k-1} \sum_{l'=1}^L H^{\rm eff}_{k-k'}(l',l)
\langle n | \psi_{k'}^{(l')} \rangle
\right),
\end{equation}
where $E_0$ is the unperturbed energy of the states in the degenerate
manifold and $E_n=\langle n | H_0 | n \rangle$.
These last two equations are the desired recurrence relations.

The next problem to address is how to carry out subgraph subtraction,
so that one can obtain an exact $n$th order expansion by considering
only a finite number of connected clusters.
(See for example Ref.~\cite{peqmbs} for a general discussion of subgraph
subtraction, which is at the heart of the cluster expansion approach
to constructing series expansions.)
But at the moment, it is not entirely obvious what quantities one is
supposed to be subtracting!
The point one has to establish is that {\it weight\/} of
the quantity $Q$ under consideration vanishes for a disconnected
graph; or, equivalently, for a graph $C$ consisting of
two unconnected clusters $A$ and $B$, $Q_C=Q_A + Q_B$.

Since we are going to have to discuss specific quantities it
is advantageous to consider a specific model, and for the
sake of clarity we choose a particularly simple one: the $S=1/2$
transverse-field Ising model defined by
\begin{equation}
H = -\sum_i \sigma^z_i
-\lambda\sum_{\langle ij \rangle} \sigma^x_i \sigma^x_j .
\end{equation}
We will consider a ``disordered state'' expansion, in which
$H_0$ is the first term and $H_1$ the second term (except for
the factor $\lambda$).  Although the explicit discussion
in the  next several paragraphs will be concerned entirely with
this model, it should be fairly clear that other models
(such as the Kondo lattice model, variants of the $t$-$J$ model,
the Heisenberg-Ising model, and other models for which the
ground-state properties have been studied by means of series expansions)
may be treated along similar lines.

For any $N$-site cluster the ground state of $H_0$
is the unique state with $\sigma^z_{\rm tot}=N$, namely,
\hbox{$|\uparrow\ldots(\hbox{$N$ times})\uparrow\rangle$}.
The lowest excited states form an $N$-fold degenerate set, the
most natural basis set being
\hbox{$|\downarrow\uparrow\ldots(\hbox{$N-1$ times})\uparrow\rangle$}, etc.
One can label these ``single-flipped-spin'' states by the site of
the cluster which carries the flipped spin.
In the thermodynamic limit, one could equally well say that $H_0$ has
a dispersionless mode with energy 2.
We will now proceed to show how one can use perturbation theory
to determine how this spectrum evolves as the Ising coupling is turned on.

Suppose we have a cluster $C$ with two disconnected components $A$ and $B$,
as described above.  (We will now attach $A$, $B$, or $C$ as subscripts
to a variety of quantities, to indicate that they are associated with
one of these clusters).  Consider the structure of $H^{\rm eff}_C$
for the single-flipped-spin states.
At order $\lambda^0$, the states on which $H^{\rm eff}_C$
acts have a flipped spin in one of either $A$ or $B$ and no flipped spin
in the other component, and it should be
clear that even as $\lambda$ is increased from zero, and the states
develop more structure, that there is no way for
the spin-flip to ``jump'' between disconnected parts of the cluster.
In other words, we assert that if we start with an unperturbed state in
which there is a spin-flip in subcluster $A$ but no spin-flip in $B$,
and apply $H_1$  (more precisely,
the restriction of $H_1$ to $C$) any number of times, the resulting state
will have zero overlap with any unperturbed state in which there is
a spin-flip in $B$ but none in $A$.
What this implies is that
$H^{\rm eff}_C$ has a block-diagonal form, specifically,
\begin{equation}
H^{\rm eff}_{C} = [H^{\rm eff} + e_B I]_A
                \oplus
                [H^{\rm eff} + e_A I]_B
\end{equation}
where $e_A$ denotes the ground state energy of cluster $A$ and $I$ is
the identity operator.
Thus $H^{\rm eff}$ {\it itself} does not have a cluster expansion,
since it is not simply additive for a disconnected cluster.
However, if we subtract $e_C=e_A+e_B$ from all the diagonal elements
on both sides of the equation above,
\begin{equation}
[H^{\rm eff} - e_C I]_{C} = [H^{\rm eff} - e_A I]_A
                \oplus
                [H^{\rm eff} - e_B I]_B ,
\label{ABC}
\end{equation}
we see that $H^{\rm eff}-e I$ {\it does\/} have a cluster expansion.
In carrying out subgraph subtraction, one must keep track of which sites
in the graph correspond to which sites in the subgraph, and the subtract
all of the appropriate matrix elements, order by order.
As a check of the validity of the calculation, one should find
that the weight for $H^{\rm eff}-e I$
on a cluster with $p$ terms
of $H_1$ should vanish identically up to order $\lambda^{p-1}$.

Once the weights have been evaluated for all graphs contributing
up to a desired order in $\lambda$,
the excited state dispersion is trivial to evaluate.
Each matrix element in the weight of $H^{\rm eff}-eI$ for each cluster
is associated with some vector on the lattice. One must simply sum
all of the matrix elements associated with any given lattice vector
${\bf r}_s$; denote the resulting sum of (power series of) matrix elements
by $t_s$.  The effective Hamiltonian in the single-flipped-spin sector
{\it for the infinite lattice\/} is now known to the desired order in
$\lambda$, and it is diagonalized by plane-wave
eigenfunctions.
The resulting dispersion relation for the excitation energy associated
with a spin-flip is  $\sum_s t_s \exp (i {\bf q} \cdot {\bf r}_s)$.

Just as in the case of ground-state expansions, one should take advantage
of geometric and topological classification of the clusters in order to
avoid redundant calculations:  but that is a well-understood
technical detail that requires no further discussion here.

We have in fact implemented the algorithm described above for
the $S=1/2$ transverse-field
Ising chain, and reproduced the exactly-known dispersion \cite{pfeuty}
$[1+2\lambda\cos q + \lambda^2]^{1/2}$ to order $\lambda^{11}$
(and for this model it would be easy to push the calculations further).
Upon being informed of the algorithm,
Singh \cite{singhunpub} promptly carried out the analogous calculation
for the single-flipped-spin excitations of the $S=1/2$ square-lattice
Heisenberg-Ising antiferromagnet; by appropriate extrapolation to
the Heisenberg limit he directly
obtains a value of the spin-wave velocity which is consistent
with the current best estimates.

Similar calculations of elementary
excitation spectra for other models would be straightforward;
however, they do not exhaust the potential applications of this method.
The ``eigenvectors'' $|\psi^{(l)}\rangle$ contain
information on correlations in the vicinity of the excitation.
The set of overlaps (in the context of the transverse-field Ising model,
for concreteness) $\langle \psi | \sigma^x_{\l'} | \psi^{(l)}\rangle$,
where $|\psi\rangle$ is the perturbatively-constructed ground state of
the cluster, should contain all the information needed to evaluate
the quasiparticle residues associated with the elementary excitations.
In fact a calculation of the residues has already been accomplished for
the Heisenberg-Ising antiferromagnet, and will be reported
separately.  One subtle point worth emphasizing here is that the method
of constructing $H^{\rm eff}$ is a similarity transformation
but not a unitary transformation.  Insofar as the calculation of
the spectrum is concerned, the distinction is irrelevant.  However,
the states $ | \psi^{(l)}\rangle$ are not orthogonal, and this fact must
be carefully taken into account whenever they are used in further
calculations.

Another interesting extension would be to ``two-particle'' excited
states.  Here the situation is rather more complicated than
for ``single-particle'' excitations such as the one-flipped-spin
states of the transverse field Ising model, on two counts.
First, the two-particle effective Hamiltonian for
the infinite lattice is not, in general, trivially diagonalizable like
in the one-particle case.  So the value of this
method would be to yield a finite number of terms in
an exact series expansion for the
quasiparticle interactions:  the next step would be along the
lines of traditional many-body theory, or perhaps even exact
diagonalization!  (One would have dramatically reduced the
size of the Hilbert space, compared to exactly diagonalization of
the original model, but at the cost of introducing more
complicated, long-range interactions.)  Second, the analog
of Eq.~(\ref{ABC}) is considerably more difficult to write down,
and it is not clear precisely what quantity exhibits a cluster expansion.
We believe these issues merit examination in future work, as well.

At this point it is possible to sensibly compare the method outlined
here with the few other series expansions methods for excited states
which exist in the literature.  Expansions for the elementary
excitation spectra of the transverse-field Ising model on
various lattices, and of the dimerized Heisenberg antiferromagnetic
chain, were carried out some years ago by
Pfeuty and Elliott \cite{pfeutyelliott} and Harris \cite{harris},
respectively.  Neither of these methods is
in the form of an explicit cluster expansion, nor do they appear to be
readily implementable as a computer program, as required for
high-order calculations.  So far as we are aware these methods
have not seen further use.

Barber and Duxbury \cite{barberduxbury}
constructed the expansion for the lowest
excitation energy in the next-neighbor transverse field Ising chain
via Rayleigh-Schr\"odinger perturbation theory on finite rings.
Because this is not a cluster expansion method it cannot be efficiently
extended to higher-dimensional lattices, and it is not clear whether
it can be adapted so as to yield the entire elementary excitation
spectrum, either.
However, this approach has been applied to other one-dimensional
models (see for example \cite{hornbybarber}).

Finally, there is a method which is due to B. G. Nickel
\cite{nickel} and which has been applied by Hamer and coworkers
to several spin models \cite{Heetal,hamerirving}.
This method is {\it almost\/} a cluster expansion; to be precise
a finite set of disconnected clusters need to be included
as well as the connected clusters.
It is also limited to the lowest excitation energy only.

What sets the method presented in this paper apart from those already
reported in the literature is the recognition that there {\it is\/}
an excited state quantity which is additive over disconnected clusters,
so that the usual machinery of connected-cluster expansions can
be applied.  Because this quantity turns out to be the effective
Hamiltonian in the single-particle excitation subspace, one can obtain
the entire elementary excitation spectrum as readily as the gap.

{\it Acknowledgements.}
I would like to thank Rajiv Singh for many discussions, and
for bringing Ref.~\cite{hamerirving} to my attention.
The seeds of this work were planted during
conversations with Stuart Trugman in the course of a
visit to Los Alamos National Laboratory which was supported by
Associated Western Universities; I am grateful for the hospitality
of Kevin Bedell during that visit.
This work is supported by NSF Grant DMR 94--57928.

\end{document}